\pdfoutput=1
\documentclass[11pt,twoside,a4paper,cmspaper,final,collab]{cms-tdr}
\def\svnVersion{0d2a356}\def\svnDate{2025/06/12}\def\cmsCernNoTag{CERN-EP-2024-327}\def\cmsCernDate{\today}\def\cmsMessage{Published in Physics Letters B as \href{http://dx.doi.org/10.1016/j.physletb.2025.139651}{\doi{10.1016/j.physletb.2025.139651}.}}
\begin{document}\cmsNoteHeader{SMP-24-007}

\newcommand{\PQuv}{{\HepParticle{\PQu}{\mathrm{v}}{}}\xspace}
\newcommand{\PQdv}{{\HepParticle{\PQd}{\mathrm{v}}{}}\xspace}

\newcommand{\as}{\ensuremath{\alpS}\xspace}
\newcommand{\asmz}{\ensuremath{\as(m_{\PZ})}\xspace}
\newcommand{\asq}{\ensuremath{\as(Q)}\xspace}
\newcommand{\asmur}{\ensuremath{\as(\mur)}\xspace}
\newcommand{\mur}{\ensuremath{\mu_{\mathrm{r}}}\xspace}
\newcommand{\muf}{\ensuremath{\mu_{\mathrm{f}}}\xspace}
\newcommand{\mt}{\ensuremath{m_{\PQt}}\xspace}
\newcommand{\mb}{\ensuremath{m_{\PQb}}\xspace}
\newcommand{\mc}{\ensuremath{m_{\PQc}}\xspace}
\newcommand{\fs}{\ensuremath{f_{\PQs}}\xspace}
\newcommand{\sqrts}{\ensuremath{\sqrt{s}}\xspace}
\newcommand{\pp}{\ensuremath{\Pp\Pp}\xspace}
\newcommand{\chisq}{\ensuremath{\chi^2}\xspace}
\newcommand{\Ndof}{\ensuremath{N_{\mathrm{dof}}}\xspace}
\newcommand{\Ndp}{\ensuremath{N_{\mathrm{p}}}\xspace}
\newcommand{\absrap}{\ensuremath{\abs{y}}\xspace}

\newcommand{\xFitter}{\ensuremath{\textsc{xFitter}}\xspace}
\newcommand{\AbsoluteScale}{\ensuremath{\textit{AbsoluteScale}}\xspace}
\newcommand{\AbsoluteMPFBias}{\ensuremath{\textit{AbsoluteMPFBias}}\xspace}
\newcommand{\SinglePionECAL}{\ensuremath{\textit{SinglePionECAL}}\xspace}
\newcommand{\SinglePionHCAL}{\ensuremath{\textit{SinglePionHCAL}}\xspace}
\newcommand{\Fragmentation}{\ensuremath{\textit{Fragmentation}}\xspace}
\newcommand{\RelativeJER}{\ensuremath{\textit{RelativeJER}}\xspace}
\newcommand{\RelativePt}{\ensuremath{\textit{RelativePt}}\xspace}
\newcommand{\RelativeFSR}{\ensuremath{\textit{RelativeFSR}}\xspace}
\newcommand{\RelativeStat}{\ensuremath{\textit{RelativeStat}}\xspace}
\newcommand{\FlavourQCD}{\ensuremath{\textit{FlavourQCD}}\xspace}
\newcommand{\TimePt}{\ensuremath{\textit{TimePt}}\xspace}
\newcommand{\TimeEta}{\ensuremath{\textit{TimeEta}}\xspace}

\newlength\cmsTabSkip\setlength{\cmsTabSkip}{1ex}
\ifthenelse{\boolean{cms@external}}{%
    \providecommand{\cmsTable}[1]{\resizebox{\linewidth}{!}{#1}}
}{%
    \providecommand{\cmsTable}[1]{#1}
}
\ifthenelse{\boolean{cms@external}}{\newcommand{\cmsApp}{}}{\newcommand{\cmsApp}{Appendix}}

\cmsNoteHeader{SMP-24-007}
\title{Determination of the strong coupling and its running from measurements of inclusive jet production}
\date{\today}
\abstract{The value of the strong coupling \as is determined in a comprehensive analysis at next-to-next-to-leading order accuracy in quantum chromodynamics. The analysis uses double-differential cross section measurements from the CMS Collaboration at the CERN LHC of inclusive jet production in proton-proton collisions at centre-of-mass energies of 2.76, 7, 8, and 13\TeV, combined with inclusive deep-inelastic data from HERA. The value $\asmz=0.1176\,^{+0.0014}_{-0.0016}$ is obtained at the scale of the \PZ boson mass. By using the measurements in different intervals of jet transverse momentum, the running of \as is probed for energies between 100 and 1600\GeV.}

\hypersetup{%
pdfauthor={CMS Collaboration},%
pdftitle={Determination of the strong coupling and its running from measurements of inclusive jet production},%
pdfsubject={CMS},%
pdfkeywords={CMS, strong coupling, QCD analysis, inclusive jet production}}

\maketitle

\section{Introduction}

Collimated sprays of particles, conventionally called jets,
are abundantly produced in high-energy proton-proton (\pp)
collisions at the CERN LHC.
At sufficiently high transverse momenta (\pt), jet production is described by quantum chromodynamics (QCD)
using perturbative techniques (pQCD),
which have now reached
next-to-next-to-leading order (NNLO) accuracy.

Predictions of jet production in hadronic collisions depend on the value of the strong coupling \as
and an accurate understanding of the proton structure,
encoded in the parton distribution functions (PDFs).
Precise measurements of jet production provide a means to determine
the strong coupling constant at the mass of the \PZ boson, \asmz, and the PDFs up to high values of $x$, which is the proton momentum fraction carried by the struck parton.

Inclusive jet production, $\pp\to\text{jet}+\PX$,
refers to events with at least one jet with \pt above a certain threshold
in the final state and is a key process to test pQCD calculations up to the highest accessible energy scales.

{\tolerance=800
In \pp collisions at the LHC, inclusive jet production has been extensively
studied by the
CMS~\cite{CMS:2015jdl, CMS:2012ftr,CMS:2014qtp, CMS:2016lna, CMS:2021yzl, CMS:5.02TeV} and
ATLAS~\cite{ATLAS:2010dmq, ATLAS:2011juz, ATLAS:2013pbc, ATLAS:2014riz, ATLAS:2017kux, ATLAS:2017ble}
Collaborations at several centre-of-mass energies over a wide kinematic range.
\par}

This Letter presents a determination of \asmz simultaneously obtained with the PDFs in a comprehensive QCD analysis at NNLO, where
the CMS measurements of inclusive jet production in \pp collisions at $\sqrts=2.76$~\cite{CMS:2015jdl}, 7~\cite{CMS:2012ftr,CMS:2014qtp}, 8~\cite{CMS:2016lna}, and 13\TeV~\cite{CMS:2021yzl} are used together with cross section measurements of inclusive deep-inelastic scattering (DIS) at the DESY HERA~\cite{H1:2015ubc}.
The simultaneous extraction of PDFs and \asmz mitigates their correlation and avoids a potential bias in the determination of \asmz~\cite{Forte:2020pyp}. Furthermore, the running of \as as a function of the energy scale is demonstrated at NNLO up to 1.6\TeV.

\section{The CMS detector}

The CMS apparatus~\cite{CMS:2008xjf,CMS:2023gfb} is a multipurpose, nearly hermetic detector, designed to trigger on~\cite{CMS:2020cmk,CMS:2016ngn,CMS:TRG-19-001} and identify electrons, muons, photons, and (charged and neutral) hadrons~\cite{CMS:2020uim,CMS:2018rym,CMS:2014pgm}. A particle-flow (PF) algorithm~\cite{CMS:2017yfk} aims to reconstruct all individual particles in an event, combining information provided by the all-silicon inner tracker, the crystal electromagnetic (ECAL) and the brass-scintillator hadron calorimeters (HCAL) that operate inside a 3.8\unit{T} superconducting solenoid, with data from the gas-ionization muon detectors embedded in the flux-return yoke outside the solenoid. The reconstructed PF particles are used to build \PGt leptons, jets, and missing transverse momentum~\cite{CMS:2018jrd,CMS:2016lmd,CMS:2019ctu}.

\section{Inclusive jet measurements}

In this work, four CMS measurements of the double-differential inclusive jet cross section as functions of the individual jet \pt and absolute rapidity \absrap are included.
These measurements use jets clustered with the anti-\kt~\cite{Cacciari:2008gp} algorithm as implemented in the \textsc{FastJet}~\cite{Cacciari:2011ma} program with a distance parameter of $R=0.7$.
For each of the measurements, the corresponding integrated luminosity \lumi, the number of data points \Ndp, and the ranges in \pt and \absrap are listed in Table~\ref{tab:measurements}.
Although the measurement at \(\sqrt{s} = 8\)\TeV extended up to 2.5\TeV in \(\pt\), the range from 1784 to 2500\GeV was excluded from the fit in the original analysis~\cite{CMS:2016lna} due to very large statistical correlations in these bins.

\begin{table}[!ht]
\centering
\topcaption{The CMS inclusive jet measurements used in this analysis. The columns show the centre-of-mass energy, the integrated luminosity, the number of measured data points, the ranges of individual jet \pt and \absrap, and the reference to the corresponding publication.}
\label{tab:measurements}
\renewcommand{\arraystretch}{1.1}
\cmsTable{\begin{tabular}{clcccc}
    \sqrts [{\TeVns}] & \lumi [{\!\fbinv}] & \Ndp & \pt [{\GeVns}] & \absrap &  Ref. \\[2pt] \hline
    2.76   & 0.0054 & 80   & 74--592    & 0.0--3.0  &  \cite{CMS:2015jdl}            \\
    7      & 5.0    & 130  & 114--2116  & 0.0--2.5  &  \cite{CMS:2012ftr,CMS:2014qtp} \\
    8      & 20     & 165  & 74--1784   & 0.0--3.0  &  \cite{CMS:2016lna}       \\
    13     & 33.5   & 78   & 97--3103   & 0.0--2.0  &  \cite{CMS:2021yzl}          \\
\end{tabular}}
\end{table}

The primary source of experimental systematic uncertainty in jet measurements comes from the calibration of the jet energy scale (JES) and the jet energy resolution (JER).
These calibrations and their associated uncertainties are determined using
test beam data, Monte Carlo (MC) simulations, and measurements of dijet, \PZ{}+jet, \PGg{}+jet, and multijet production~\cite{CMS:2016lmd, collaboration_2011}. We estimate the correlations between these uncertainties across \pt and \absrap intervals within individual measurements and
among the measurements at various \sqrts, and include them in the QCD analysis.
This analysis results in a recommendation for the use of these correlations in future global QCD analyses of CMS inclusive jet measurements.

The JES uncertainties are grouped into eight main categories.
A detailed description of the JES calibration and the corresponding uncertainties is given in Refs.~\cite{CMS:2016lmd, collaboration_2011}.
The naming scheme of the uncertainty sources follows from
previous publications~\cite{CMS:2015jdl,CMS:2012ftr,CMS:2014qtp,CMS:2016lna,CMS:2021yzl} and is identical to the implementation in \xFitter~\cite{Alekhin:2014irh,xFitterDevelopersTeam:2017xal}.
Additional uncertainties specific to each individual measurement are discussed
in the original publications.
\begin{itemize}
{\tolerance=800
\item Uncertainties related to the absolute scale are associated with the JES correction within the barrel region, corresponding to pseudorapidities $\abs{\eta}<1.3$.
This correction (\AbsoluteScale) is obtained from a global fit using \PZ{}+jet, \PGg{}+jet, and multijet data.
It includes a correction for initial-state (ISR) and final-state radiation (FSR) (\AbsoluteMPFBias).
Extrapolations to high \pt beyond the reach of the measurements are performed by using MC simulations.
Additional uncertainties are introduced to address response variations from different MC event generators (\Fragmentation) and the single-particle response in the ECAL and HCAL (\SinglePionECAL, \SinglePionHCAL).
\par}

\item Relative JER uncertainties represent the $\eta$-dependent uncertainty of the JES from the JER corrections (\RelativeJER).

\item The relative $\eta$ correction of the JES calibration ensures a uniform detector response across different detector regions. In addition, it includes a log-linear \pt dependence. An uncertainty associated with the choice of the \pt-dependent shape is included. (\RelativePt).

\item The relative contribution \RelativeFSR corresponds to an uncertainty in the $\eta$-depen\-dent corrections for ISR and FSR.

\item The statistical uncertainties in the determination of the relative corrections in various regions of $\eta$ for multiple uncertainty sources are merged into one uncertainty (\RelativeStat).

\item Differences in the simulated detector response to jets from different quark flavours and gluons result in the \FlavourQCD uncertainty. It is derived from differences between data and MC simulations when the corresponding corrections are applied to various mixtures of jet flavours. The uncertainty is based on response differences to jets from $\PQu\PQd\PQs$/{\PQc}/{\PQb} quarks and gluons between \PYTHIA~\cite{Sjostrand:2014zea} and \HERWIG~\cite{Bahr:2008pv}.

\item Two time-dependent uncertainties address the JES variation over time during the data-taking periods (\TimePt and \TimeEta).

\item Uncertainties associated with the corrections for additional \pp interactions within the same or nearby bunch crossings (pileup) are included.
\end{itemize}
The correlation scheme for JES-related uncertainties across the different CMS measurements used in this analysis is described below and illustrated in \cmsApp~\ref{sec:JES}.
The uncertainties in pileup, the relative statistical contribution \RelativeStat, and the time-dependent sources \TimePt and \TimeEta are considered uncorrelated across the measurements.
For the measurement at $\sqrts=7\TeV$~\cite{CMS:2012ftr}, the correlation scheme for JES-related uncertainties is used as described in Ref.~\cite{CMS:2014qtp}.
In particular, the \SinglePionECAL and \SinglePionHCAL sources are decorrelated as a function of $\eta$.
The correlations in the remaining JES uncertainties between different data sets are implemented as follows.
\begin{itemize}
\item The \AbsoluteMPFBias uncertainty  is fully correlated between the 2.76, 8, and 13\TeV measurements. In the 7\TeV analysis, a combined source of uncertainty is used by including uncertainties from the \AbsoluteScale, ISR and FSR, and a statistical component associated with the \AbsoluteScale correction~\cite{CMS:2014qtp}. This combined source of uncertainty in the 7\TeV analysis is uncorrelated with uncertainties in the other measurements.

\item The \AbsoluteScale, \SinglePionECAL, \SinglePionHCAL, and \RelativeJER uncertainties are fully correlated between the 2.76 and 8\TeV measurements, since the corresponding corrections were obtained using 8\TeV data and applied to both measurements.
These uncertainties are uncorrelated with other data sets, for which the corrections were specifically derived.

\item The \RelativeFSR and \Fragmentation uncertainties are fully correlated between the 2.76, 7, and 8\TeV measurements, but uncorrelated with the 13\TeV measurement, because of the use of different MC tunes and event generators.

\item The \RelativePt uncertainties are uncorrelated across the measurements,
since these account for the differences between linear and logarithmic fits in the \pt extrapolation of the residual corrections, which vary across data-taking periods.

\item The \FlavourQCD uncertainty is treated as fully correlated between the 7 and 8\TeV measurements because these used identical uncertainties, but uncorrelated with the 2.76 and 13\TeV ones to account for differences in simulation setup, including centre-of-mass energy, pileup, and MC tunes. 
\end{itemize}
The JER uncertainties are treated as uncorrelated across measurements at different energies. Furthermore, within each individual measurement, the JER uncertainties have been decorrelated across \absrap for the 2.76, 7, and 8\TeV measurements. This approach is motivated by the unaccounted residual \absrap dependence in these data sets. The JER uncertainty is considered correlated among \absrap and \pt for the 13\TeV measurement~\cite{CMS:2021yzl}.
An additional uncorrelated component of 1.5\% is assigned to the JER uncertainty in the 2.76\TeV measurement to take into account a statistical contribution to the JER uncertainty.

\section{Theoretical predictions}

The NNLO pQCD calculations used in this study are performed assuming five active massless quark flavours in the leading-colour (LC) and  leading-flavour-number approximation using the \textsc{nnlojet} program~\cite{Currie:2016bfm}. The subleading colour contributions for the NNLO corrections, neglected in Ref.~\cite{Currie:2016bfm}, have been recently calculated~\cite{Chen:2022tpk} and have a very small impact on inclusive jet production with $R=0.7$.
The renormalization, \mur, and factorization, \muf, scales in the calculation are set to the individual jet \pt, following the studies of Ref.~\cite{Currie:2018xkj}.
The cross section predictions for intervals in \pt and \absrap of the respective measurements are stored in the form of interpolation grids in the \textsc{APPLfast}~\cite{Britzger:2022lbf} format,
allowing for a fast evaluation of the predictions under variations of \as, PDFs, or the \mur and \muf scales.
The numerical uncertainty in the grids is below 1\% in most intervals and increases to about 5\% in the forward region at high \pt, remaining significantly smaller than the statistical uncertainties in the measurements. However, in the forward region, predictions in a few intervals at the highest values of \pt have numerical grid  uncertainties exceeding 10\% and are, therefore, excluded from the QCD interpretation.
When the interpolation grids for this analysis were produced, the recent improvements in the merging procedure of \textsc{nnlojet} that aim to reject outliers without biasing the result in multidimensional MC integrations were not available.
To account for the remaining point-to-point fluctuations, the numerical uncertainty has been increased by a factor of two. However, its impact on the fit is negligible.

The QCD predictions are modified with a multiplicative correction for contributions of the dominant electroweak (EW) effects.
These arise from the virtual exchange of soft or collinear massive weak gauge bosons, calculated at next-to-leading-order precision~\cite{Dittmaier:2012kx}.
The contribution of EW effects becomes important at large jet \pt, reaching 11\% at the highest \pt probed in the 13\TeV measurement.
Subleading EW effects from the emissions of real weak gauge bosons have been estimated in the 13\TeV measurement to be smaller than 1\% at high \pt and are neglected.
The EW corrections for the 13\TeV measurement are taken from the original publication~\cite{CMS:2021yzl}. For the 7 and 8\TeV measurements, these corrections have been updated using more recent PDF sets.
The EW corrections are omitted for the 2.76\TeV measurement because of their negligible impact for \pt smaller than 600\GeV.

To compare the fixed-order predictions to the measured particle-level cross sections, these
are corrected for the nonperturbative (NP) effects from hadronization and the underlying event.
These NP corrections have been derived as cross section ratios from event generators with parton showers, in which NP effects are switched on and off.
These corrections are applied through \pt- and \absrap-dependent correction factors
as provided in the original publications~\cite{CMS:2021yzl,CMS:2016lna,CMS:2014qtp,CMS:2015jdl} of the individual measurements.
In each measurement, the generator tune used is the one providing the best match to the data.
The NP corrections have a size between 5 and 20\% at low \pt, with their effect decreasing with increasing \pt.
The associated uncertainties range from 1 to 4\% at low \pt, depending on \absrap, and are negligible at high \pt.
The uncertainties related to NP corrections are treated correlated in ranges of \pt and \absrap in each individual measurement and uncorrelated across the measurements.
Despite slight differences in the way the NP corrections were obtained for each measurement, the impact of the NP corrections in the \asmz and PDF determination is insignificant considering their uncertainties.
In particular, the impact of the NP corrections in the QCD analysis was studied by
performing individual fits with a subset of measurements with jet $\pt>150$, 200, and 300\GeV. In this way, the kinematic ranges with decreasing importance of NP corrections are consecutively probed. The difference in the results is negligible.
The main result is based on the full \pt spectra probed by the CMS jet data.

\section{The QCD analysis}

In the QCD analysis, the four CMS measurements of the double-differential cross sections of inclusive jet production are used
together with the combined neutral current (NC) and charged current (CC) electron-proton DIS cross sections measured at HERA~\cite{H1:2015ubc}, following the general strategy of Refs.~\cite{H1:2009pze,H1:2015ubc}.

Theoretical predictions for the DIS cross sections are calculated at fixed order in QCD at NNLO accuracy using the \textsc{qcdnum} code~\cite{Botje:2010ay},
with \mur and \muf set to the squared four-momentum transfer $Q^2$.
The contribution of massive \PQc and \PQb quarks to the DIS cross sections is treated in the Thorne--Roberts
general-mass variable-flavour number scheme~\cite{Thorne:1997ga,Thorne:2006qt,Thorne:2012az}.
The masses of \PQc and \PQb quarks are set as $\mc=1.47\pm0.06\GeV$ and $\mb=4.5\pm0.25\GeV$, respectively.
The fraction of the strange quark in the sea $\fs=x\PAQs/(x\PAQd+x\PAQs)$ is assumed to be 0.40, as in
the HERAPDF2.0 analysis~\cite{H1:2015ubc} and varied within $0.32\leq\fs\leq0.48$.
The low-$Q^2$ region in DIS, where resummation effects from small-$x$ logarithms become important, is removed from the fit
by using only the DIS measurements with $Q^2>10\GeV^2$.
This cutoff is varied by $2.5\GeV^2$ up and down.
While the central values and the variations of \mc, \mb and \fs are set as in the QCD interpretation~\cite{CMS:2021yzl} of the CMS inclusive jet measurement at \sqrts=13\TeV, the minimum $Q^2$ cut for HERA data is increased from 7.5 to $10\GeV^2$. This adjustment aligns with the original HERA publication~\cite{H1:2015ubc}, which features tensions between low-$Q^2$ ($<15\GeV^2$) and high-$Q^2$ ($>150\GeV^2$) data.
The published DIS cross sections are already corrected for the dominant EW effects.

The PDFs for the gluon, valence \PQu- and \PQd-quark, and \PAQu and \PAQd quark densities
are parametrized at a starting (evolution) scale of $Q^2_0=1.9\GeV^2$, varied by $0.2\GeV^2$~\cite{CMS:2021yzl}. The $Q^2$ dependence of the PDFs and of \as is obtained by solving the DGLAP evolution equations at NNLO in pQCD as implemented in \textsc{qcdnum}.
The functional form of the PDFs for each parton $i$ at the starting scale is
\begin{equation}
    x q_i(x) = A_i x^{B_i} (1-x)^{C_i} P_i(x),
    \label{eq:param}
\end{equation}
where the $A_i$ are normalization parameters and $P_i(x)=(1+D_ix+E_ix^2)$ are polynomials
that interpolate between the small- and large-$x$ behaviour given by the $B_i$ and $C_i$ parameters.
The $D_i$ and $E_i$ terms in the polynomial expansion
are only included following the results of an optimization procedure.
In such a ''parametrization scan``, all $D_i$ and $E_i$ parameters are set to zero and then included in the fit, one at a time.
The process is monitored through the resulting change in \chisq and the scan is terminated when no further improvement is observed.
The parameters $A_{\PQuv}$ and $A_{\PQdv}$ are determined during the fit from the valence-quark number sum rule and the value of $A_{\Pg}$ is obtained from the momentum sum rule.
In the present work, the parametrization scan led to the same parameters as the most recent NNLO QCD analysis~\cite{CMS:2021yzl} of the CMS measurement of inclusive jet production at 13\TeV, using the same assumptions for the PDF parameters. This parametrization includes 15 free parameters for the PDFs.
Additional terms in the polynomial expansion proportional to $D_i$ and $E_i$ are added to estimate the parametrization uncertainty as discussed in Section~\ref{sec:results}.

The fit is performed using the \xFitter~\cite{Alekhin:2014irh,xFitterDevelopersTeam:2017xal} program.
The values of the PDF parameters and of \as are obtained in the minimization procedure using \textsc{minuit}~\cite{James:1975dr} with a \chisq goodness-of-fit function that includes experimental uncertainties as nuisance parameters $b_{i,\text{exp}}$.
The \chisq definition is the same as defined in \xFitter~\cite{Alekhin:2014irh,xFitterDevelopersTeam:2017xal} and used in most recent analyses~\cite{CMS:2021yzl},
\ifthenelse{\boolean{cms@external}}{\begin{multline}\label{eq:prof-chi2}
    \chisq(\mathbf{T},\mathbf{b}_{\text{exp}}) = \\
    \sum_{i=1}^{N_{\text{data}}} \frac{\big[M_i - T_i(1 - \sum_j\gamma_{ij}^{\text{exp}}b_{j,\text{exp}} )\big]^2}{\delta^2_{i,\text{uncor}}T^2_i+\delta^2_{i,\text{stat}}M_iT_i(1 - \sum_j \gamma_{ij}^{\text{exp}}b_{j,\text{exp}})} \\
    + \sum_{j=1}^{N_{\text{exp.sys}}}b^2_{j,\text{exp}}.
\end{multline}}{\begin{equation}\label{eq:prof-chi2}
    \chisq(\mathbf{T},\mathbf{b}_{\text{exp}}) =
    \sum_{i=1}^{N_{\text{data}}} \frac{\big[M_i - T_i(1 - \sum_j\gamma_{ij}^{\text{exp}}b_{j,\text{exp}} )\big]^2}{\delta^2_{i,\text{uncor}}T^2_i+\delta^2_{i,\text{stat}}M_iT_i(1 - \sum_j \gamma_{ij}^{\text{exp}}b_{j,\text{exp}})}
    + \sum_{j=1}^{N_{\text{exp.sys}}}b^2_{j,\text{exp}}.
\end{equation}}
The index $i$ runs over all $N_{\text{data}}=1469$ data points of the inclusive jet and HERA measurements.
Here, $M_i$ and $T_i$ are the measurements and theory predictions, respectively. Further, $\delta_{i,\text{uncor}}$ and $\delta_{i,\text{stat}}$ denote the statistical and uncorrelated systematic uncertainties of the measurement $i$.
The experimental correlated uncertainty sources are included using the set of parameters $b_{j,\text{exp}}$, with the impact of each source on the theory point $T_i$ described by the matrices $\gamma^{\text{exp}}_{ij}$.
Asymmetric uncertainties in the PDF parameters are obtained from a $\Delta\chisq=1$ criterion using an iterative Hessian approach, following the prescription of Ref.~\cite{Pumplin:2001ct}. The uncertainties quoted in the following correspond to
68\% confidence level (\CL).

\section{Results}
\label{sec:results}

\begin{figure*}[!pt]
\centering
\includegraphics[width=0.8\textwidth]{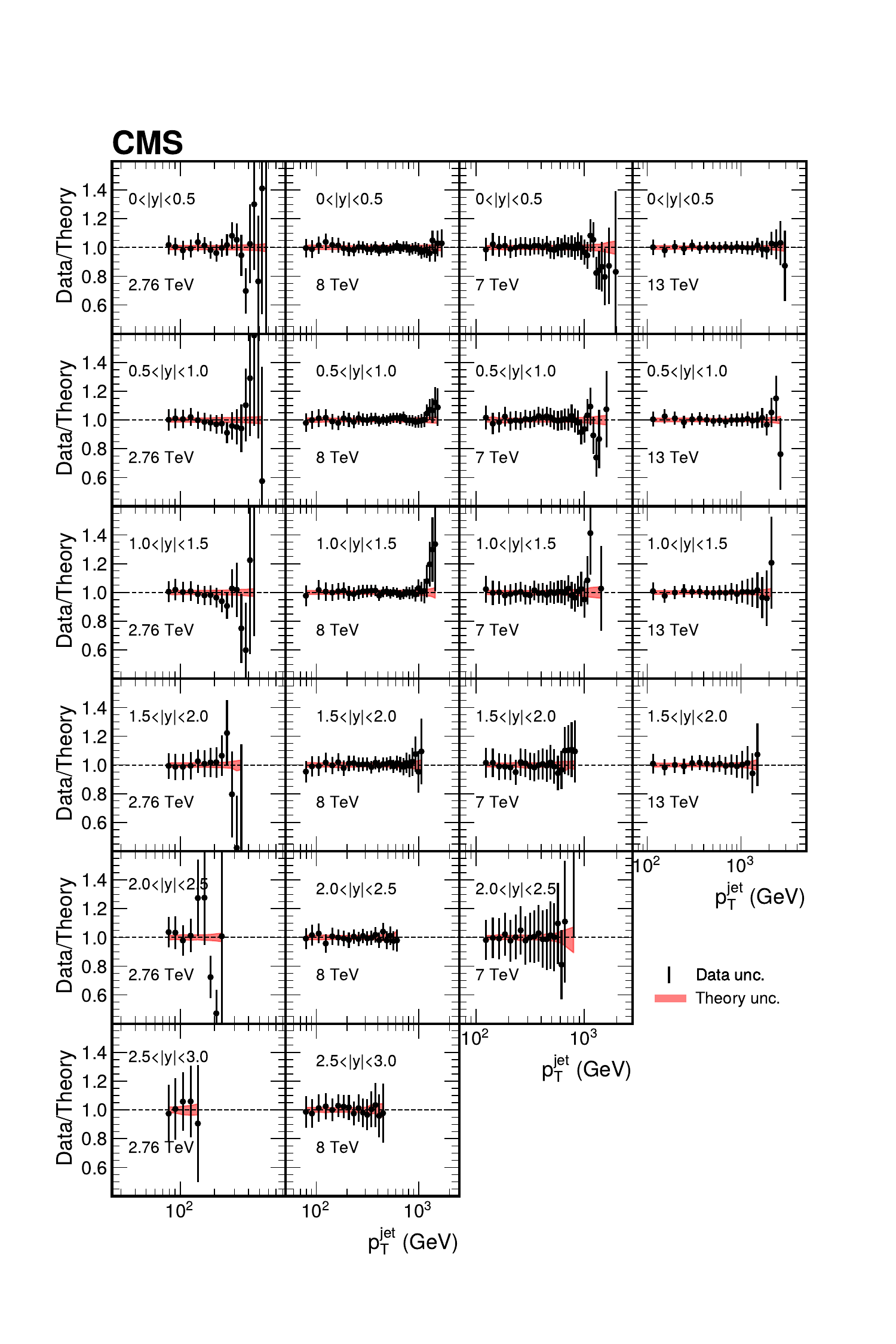}
\caption{Cross section ratios between data and theory for inclusive jet production in \pp collisions as measured by CMS at \sqrts of 2.76, 8, 7, and 13\TeV, using the anti-\kt clustering algorithm with $R=0.7$, as functions of individual jet \pt in intervals of absolute rapidity \absrap. Shown is the total uncertainty of every data point (vertical bar). Predictions correspond to the NNLO LC QCD calculations corrected for NP and EW effects. The PDFs obtained in this work are used. The total theory uncertainty (red shaded band) includes the PDF and scale variation uncertainties.}
\label{fig:Data-Th}
\end{figure*}

In Fig.~\ref{fig:Data-Th}, the measurements divided by the NNLO LC QCD predictions are shown, corrected for NP and EW effects. In general, a very good agreement between the theory predictions and the data is observed.
The consistency of all the experimental data is quantified by
the goodness of the fit \chisq from Eq.~\eqref{eq:prof-chi2}.
The values of \chisq per \Ndp are reported for the individual data sets (partial $\chisq/\Ndp$) in Table~\ref{tab:chi2}, together with the total \chisq per number of degrees of freedom \Ndof, which is $1680/1453$.
A somewhat higher partial $\chisq/\Ndp$ for the HERA DIS data is investigated in detail in the original work~\cite{H1:2015ubc}, where tensions between the measurements at low $Q^2$ (${<}15{\GeV}^2$) and high $Q^2$ (${>}150{\GeV}^2$) were identified. 
The sum of partial $\chisq/\Ndp$ for all CMS jet measurements, including their correlation, is $496/453$.

\begin{table}[!ht]
\centering
\topcaption{The values of \chisq per \Ndp for each individual data set as obtained in the fit to HERA and CMS jet data, together with the contribution to the \chisq from correlated uncertainty sources. In the last line, the total \chisq per number of degrees of freedom, \Ndof, is reported.}
\label{tab:chi2}
\renewcommand{\arraystretch}{1.1}
\begin{tabular}{lc}
    Data set &  Partial $\chisq/\Ndp$ \\[2pt] \hline
    HERA I+II neutral current  & $1036/935$ \\
    HERA I+II charged current  & $112/81$ \\
    CMS jets 2.76\TeV  & $63/80$ \\
    CMS jets 7\TeV  & $81/130$ \\
    CMS jets 8\TeV  & $206/165$ \\
    CMS jets 13\TeV  & $77/78$ \\
    Correlated \chisq HERA&  56\\
    Correlated \chisq Jets&   69\\[\cmsTabSkip]
    Total $\chisq/\Ndof$ & $1680/1453$ \\
\end{tabular}
\end{table}

\ifthenelse{\boolean{cms@external}}{\begin{figure*}[!p]}{\begin{figure*}[!t]}
\centering
\includegraphics[width=0.9\textwidth]{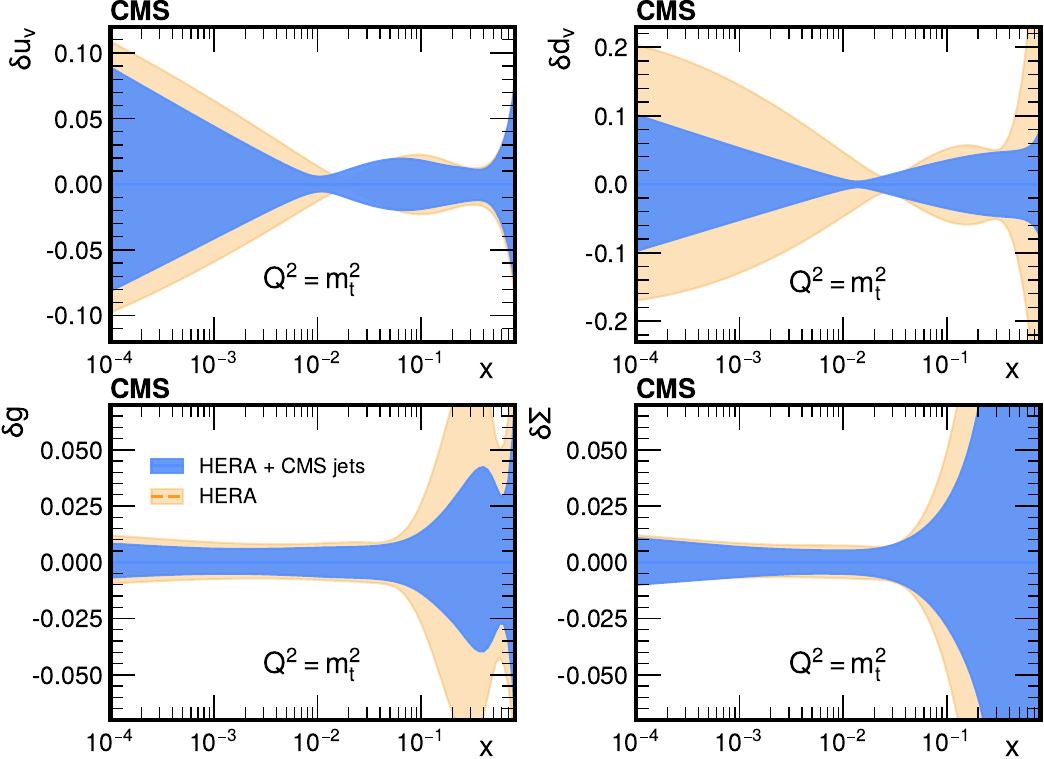}
\caption{The fit uncertainties in the valence \PQu quark (\PQuv) (upper left), valence \PQd quark (\PQdv) (upper right), gluon \Pg (lower left), and sea quark \PGS (lower right) distributions, shown as a function of $x$ at the factorization scale $Q^2=\mt^2$. The results of the HERA+CMS fit (blue shaded area) are compared with the results of the HERA-only fit (orange shaded area). In the HERA-only fit, due to poor sensitivity of the DIS data to \asmz, its value is fixed to that of the HERA+CMS fit. The uncertainties are given at 68\% \CL.}
\label{fig:pdfsHERA}
\end{figure*}

In Fig.~\ref{fig:pdfsHERA}, the impact of the CMS jet data on the PDF determination (\textit{HERA+CMS} fit) is illustrated, considering
the Hessian fit uncertainties.
The uncertainties in the PDFs resulting from the present QCD analysis are shown in comparison with the results of an alternative fit, which includes only HERA DIS data (\textit{HERA-only} fit). In the HERA-only fit, due to poor sensitivity of the DIS data to \asmz, its value is fixed to that of the HERA+CMS fit. The PDFs are shown at the factorization scale $Q^2=\mt^2$, where \mt is the top quark mass. This particular scale is chosen for enabling comparisons with previous results~\cite{CMS:2021yzl}.
Significant improvement in the uncertainties of all PDFs is observed once CMS jet measurements are included in the analysis.
Whereas improvements in the precision of the gluon PDF have been reported in earlier QCD analyses using individual inclusive jet measurements, the inclusion of all available CMS inclusive jet measurements at different \sqrts provides additional constraints on the \PQdv distribution.

In Fig.~\ref{fig:pdfs}, the PDFs obtained in the HERA+CMS fit are compared with those obtained by the global fitting groups HERAPDF20~\cite{H1:2015ubc}, NNPDF40~\cite{NNPDF:2021njg}, CT18NNLO~\cite{Hou:2019efy}, and MSHT20~\cite{Bailey:2020ooq}. These global PDFs are based on inclusive HERA DIS data and use some of the CMS jet measurements included in the present work, except for the HERAPDF20 PDF set.
In general, the PDFs determined in this analysis are in resonable agreement with the global PDF sets, particularly in the $x$ range probed by the CMS jet data.
By adding these data, the valence quark PDF, in particular the valence \PQd quark (\PQdv) distribution, is observed to agree better with the global PDFs NNPDF40 and MSHT20 than HERAPDF20.
In the HERA+CMS fit, the value of \asmz is determined together with the PDFs, resulting in $\asmz=0.1176\pm0.0009$,
where the uncertainty denotes the Hessian fit uncertainties.

\ifthenelse{\boolean{cms@external}}{\begin{figure*}[!p]}{\begin{figure*}[!h]}
\centering
\includegraphics[width=0.9\textwidth]{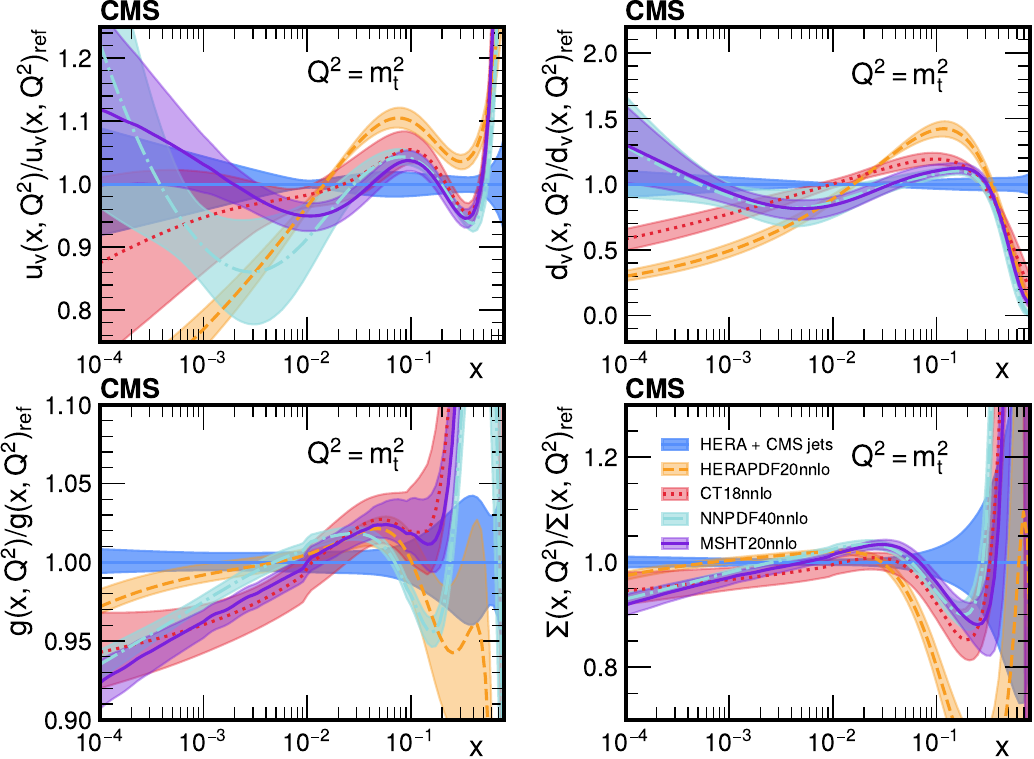}
\caption{Ratios of different global PDF sets at NNLO, namely
HERAPDF20~\cite{H1:2015ubc}, CT18~\cite{Hou:2019efy}, NNPDF4~\cite{NNPDF:2021njg}, and MSHT20~\cite{Bailey:2020ooq}, to the result of the present study.
The valence \PQu quark (\PQuv) (upper left), valence \PQd quark (\PQdv) (upper right), gluon \Pg (lower left), and sea quark \PGS (lower right) distributions are shown as functions of $x$ at the factorization scale $Q^2=\mt^2$. The Hessian PDF uncertainties at 68\% \CL are shown, where the width of the band represents the uncertainty in the numerator by keeping the denominator constant.}
\label{fig:pdfs}
\end{figure*}

Additional uncertainties are included in the QCD analysis following the strategy of Ref.~\cite{CMS:2021yzl}.
These uncertainties are the model uncertainty, obtained from variations of the fit input parameters, 
such as \mb, \mc, \fs, the minimum $Q^2$ for DIS data $Q^2_0$, the uncertainty originating from missing higher-order corrections, 
and the parametrization uncertainty, accounting for alternative PDF parametrizations.
For each variation, an alternative fit is performed. The contributions to the model uncertainty are added in quadrature, resulting in $\delta\asmz={}^{+0.0006}_{-0.0004}\,\text{(model)}$.
The uncertainty from missing higher orders in the calculation of the jet cross sections is evaluated
by varying the scales \mur and \muf (scale uncertainty) independently by a factor of two up and down, avoiding cases with $\muf/\mur=4$ or $1/4$.
For each scale choice, an individual fit is performed, and the maximum difference to the central result is included as an uncertainty of
$\delta\asmz={}^{+0.0009}_{-0.0012}\,\text{(scale)}$.
The parametrization uncertainty is estimated by extending the functional form of the PDFs by additional parameters $D$ and $E$, added one at a time. An uncertainty $\delta\asmz={}^{+0}_{-0.00004}\,\text{(param.)}$ is obtained from an envelope of the results of the corresponding fits and
is added linearly to the aforementioned uncertainties.

Summarizing the findings above, the value of \asmz is obtained as
\ifthenelse{\boolean{cms@external}}{\begin{multline}\label{NNLO_alphas}
    \asmz = 0.1176
    \,^{+0.0009}_{-0.0009}\,\text{(fit)}
    \,^{+0.0006}_{-0.0004}\,\text{(model)} \\
    \,^{+0.0009}_{-0.0012}\,\text{(scale)}
    \,^{+0.}_{<(-0.0001)}\,\text{(param.)},
\end{multline}}{\begin{equation}\label{NNLO_alphas}
    \asmz = 0.1176
    \,^{+0.0009}_{-0.0009}\,\text{(fit)}
    \,^{+0.0006}_{-0.0004}\,\text{(model)}
    \,^{+0.0009}_{-0.0012}\,\text{(scale)}
    \,^{+0.}_{<(-0.0001)}\,\text{(param.)},
\end{equation}}
corresponding to the total uncertainty of $^{+0.0014}_{-0.0016}\,\text{(tot)}$.
This value is in good agreement with the PDG world average of $\asmz=0.1180\pm0.0009$~\cite{ParticleDataGroup:2024cfk} and with previous CMS results obtained at NNLO~\cite{CMS:2021yzl,CMS:2023fix,CMS:2018fks, CMS:2014rml, CMS:2019oeb}, as shown in Fig.~\ref{fig:summaryplot}.
The inclusion of data from multiple centre-of-mass energies improves the precision by significantly reducing the fit uncertainty. The dominant contribution to the uncertainty originates from missing higher order corrections.

\begin{figure}[!htp]
\centering
\ifthenelse{\boolean{cms@external}}{%
    \includegraphics[width=\linewidth]{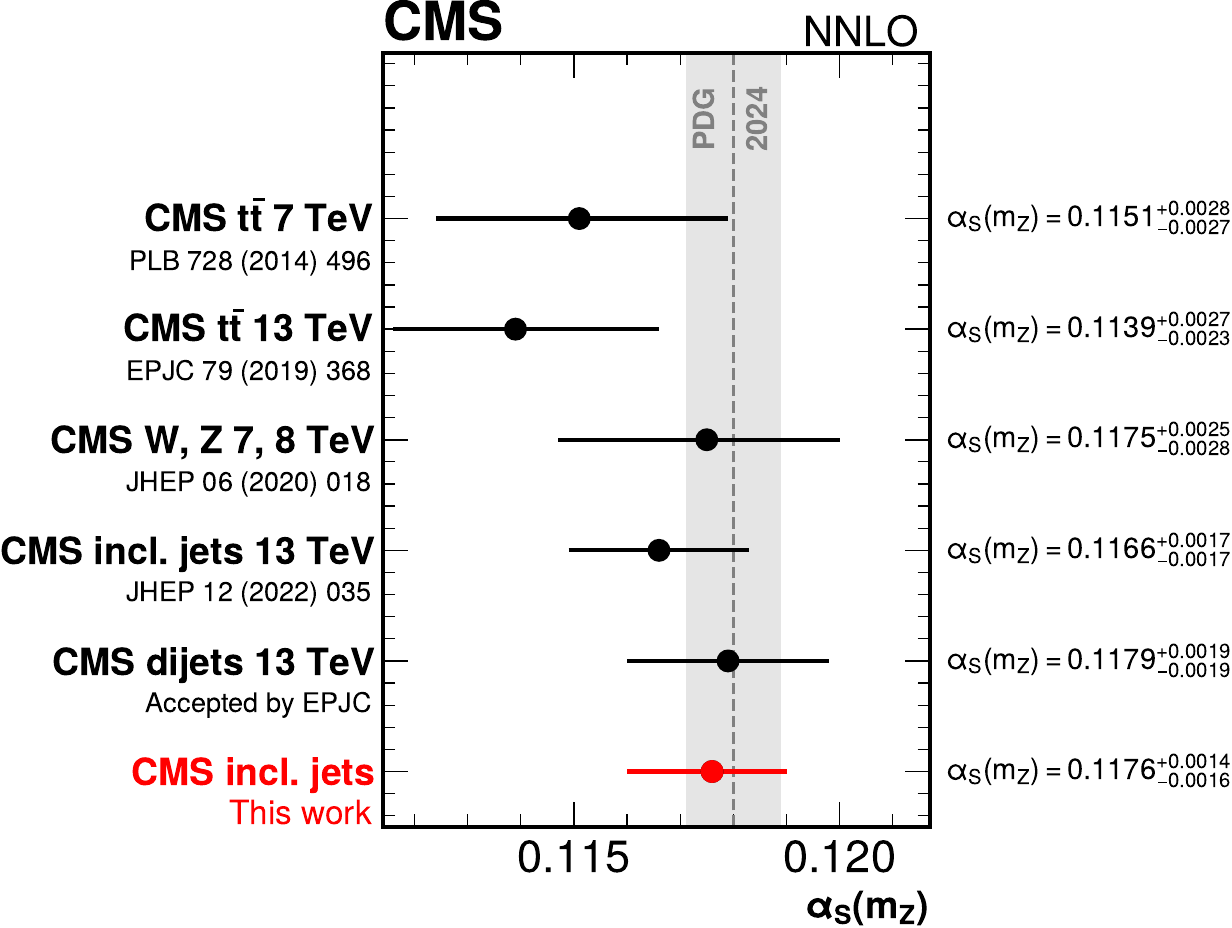}%
}{%
    \includegraphics[width=0.75\textwidth]{Figure_004.pdf}%
}
\caption{The value of \asmz obtained in this analysis (red marker), compared with
all CMS results obtained at NNLO by using different methods (black markers) with their total uncertainties (horizontal error bars).
The PDG world average (dashed line) together with its uncertainty (shaded band) is also shown.}
\label{fig:summaryplot}
\end{figure}

The value of \as in five different ranges of $\mur=\pt$ is determined, following the previous approach, which illustrates the running of the strong coupling, \asmur at NNLO. For this purpose, the CMS measurements of inclusive jet production are split into exclusive ranges of individual jet \pt.
For each \pt range, a simultaneous fit of PDFs and \asmz is performed individually.
The values of \asmz obtained in each individual fit are evolved as \asmur using the five-loop five-flavour renormalization group equation (RGE) encoded in \textsc{CRunDec}~\cite{Herren:2017osy} version 0.5.2. The same RGE is used to obtain the corresponding uncertainties. The fit, model, scale, and parametrization uncertainties are computed as in the case of Eq.~\eqref{NNLO_alphas}.
In each \pt range, \mur is calculated at NNLO QCD as a cross section-weighted average $\langle\pt\rangle$.
Results are summarised in Table~\ref{tab:running} and shown in Fig.~\ref{fig:summaryplotRunning}, where the values of \asmur are compared with the evolution of the world average of \asmz performed at five-loop order in QCD using \textsc{CRunDec}.

\begin{table*}[!ht]
\centering
\topcaption{The determined \asmz and the corresponding \asq values for each \pt range with their total uncertainties. For \asq, the individual uncertainty contributions (fit, scale, model, parametrization) are listed. Values reported as 0.0 indicate uncertainties smaller than 0.0001, while ''--`` indicates that the corresponding upper or lower uncertainty does not exist since it is one-sided.}
\renewcommand{\arraystretch}{1.4}
\begin{tabular}{ccccccccc}
\pt (GeV) & $\langle Q\rangle$ & \asmz (tot) & \asq & (fit) & (scale) & (model) & (param.) & (tot) \\ \hline
74--220  & 103.06 & 0.1182 $^{+0.0014}_{-0.0013}$ & 0.1160 & $^{+0.0011}_{-0.0011}$ & $^{+0.0007}_{-0.0005}$ & $^{+0.0006}_{-0.0004}$ & $^{+0.0}_{\hphantom{00}-}$ & $^{+0.0014}_{-0.0012}$ \\
220--395  &  266.63 & 0.1184 $^{+0.0011}_{-0.0012}$ & 0.1019 & $^{+0.0007}_{-0.0007}$ & $^{+0.0}_{-0.0004}$ & $^{+0.0004}_{-0.0003}$ & $^{\hphantom{00}-}_{-0.0}$ & $^{+0.0008}_{-0.0009}$\\
395--638  & 464.31 & 0.1179 $^{+0.0013}_{-0.0012}$&  0.0947  & $^{+0.0007}_{-0.0007}$ & $^{+0.0003}_{-0.0}$ & $^{+0.0004}_{-0.0003}$ & $^{\hphantom{00}-}_{-0.0}$ & $^{+0.0009}_{-0.0008}$\\
638--1410  & 753.66 & 0.1184 $^{+0.0014}_{-0.0012}$&  0.0898 & $^{+0.0006}_{-0.0006}$ & $^{+0.0003}_{-0.0}$ & $^{+0.0004}_{-0.0003}$ & $^{+0.0}_{\hphantom{00}-}$ & $^{+0.0008}_{-0.0007}$\\
1410--3103  & 1600.5 & 0.1170 $^{+0.0021}_{-0.0016}$& 0.0821 & $^{+0.0007}_{-0.0007}$ & $^{+0.0004}_{\hphantom{00}-}$ & $^{+0.0005}_{-0.0003}$ & $^{+0.0}_{\hphantom{00}-}$ & $^{+0.0010}_{-0.0008}$
\end{tabular}
\label{tab:running}
\end{table*}

\begin{figure}[!hb]
\centering
\ifthenelse{\boolean{cms@external}}{%
    \includegraphics[width=\linewidth]{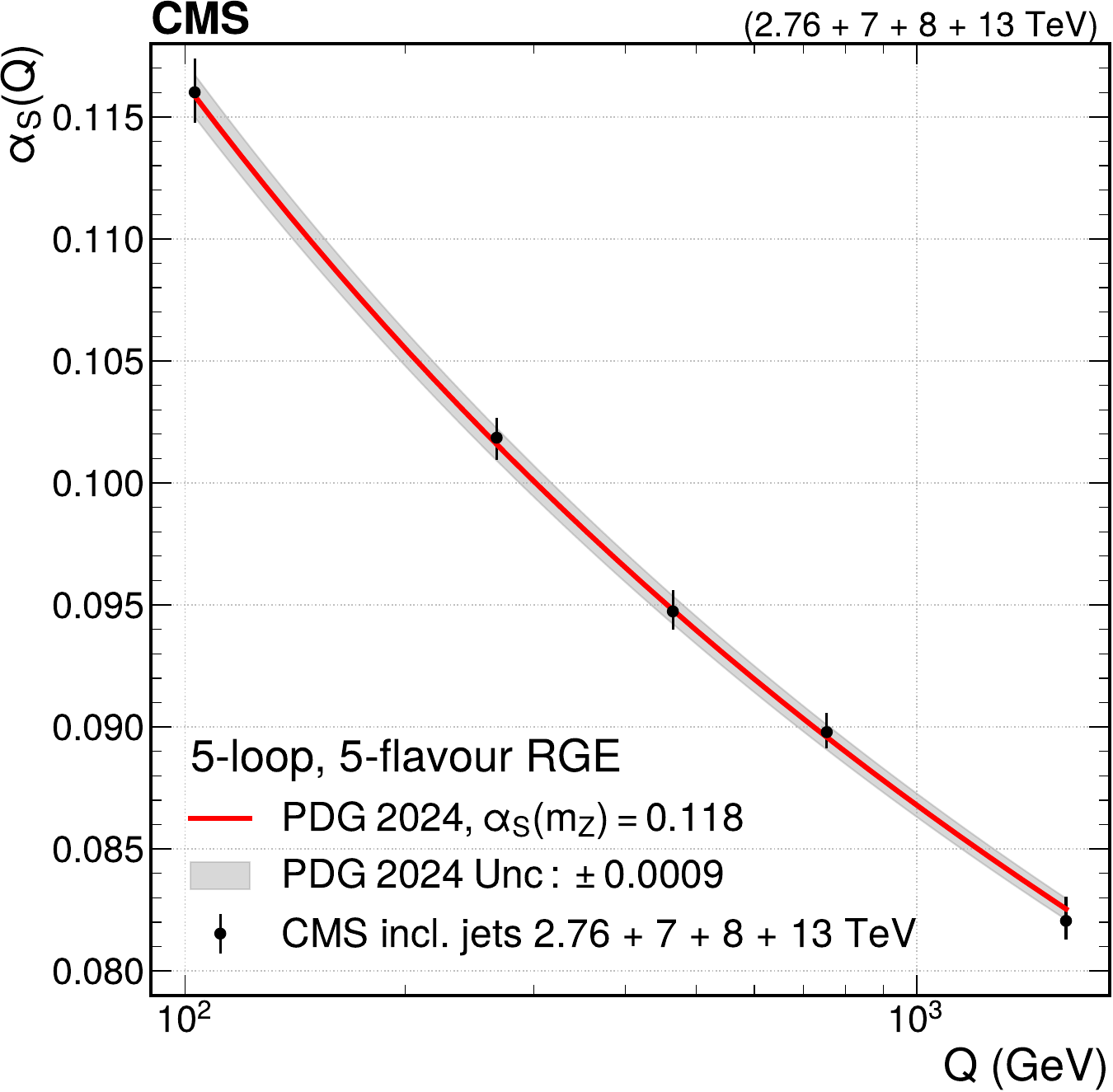}%
}{%
    \includegraphics[width=0.6\textwidth]{Figure_005.pdf}%
}
\caption{Values of \as as extracted from different jet transverse momentum ranges in the present QCD analysis at NNLO, each translated to a single scale $\langle Q\rangle$, as indicated in Table~\ref{tab:running}.
The results (black markers) are shown with their total uncertainties (vertical error bars). For comparison, the RGE at 5 loops is shown using the current world-average value $\asmz=0.1180\pm0.0009$~\cite{ParticleDataGroup:2024cfk} (red line) together with its associated total uncertainty (shaded band).}
\label{fig:summaryplotRunning}
\end{figure}

\section{Summary}

The value of the strong coupling constant at the scale of the mass of the \PZ boson, \asmz, is extracted together with the parton distribution functions (PDFs) of the proton in an analysis at next-to-next-to-leading order (NNLO) accuracy in quantum chromodynamics (QCD). The simultaneous extraction of \asmz and PDFs mitigates a possible bias from their correlations. The analysis includes electroweak effects at next-to-leading order and nonperturbative corrections. The CMS measurements of inclusive jet production in proton-proton collisions at $\sqrts=2.76$, 7, 8, and 13\TeV, using the anti-\kt clustering algorithm with a distance parameter of 0.7, are used
together with the inclusive deep inelastic scattering cross sections from HERA.
The correlations among the systematic uncertainties of each individual jet cross section measurement, as well as across all centre-of-mass energies, are reevaluated.
These correlations should be included in future global QCD analyses of these data.
By using the CMS jet data in the fit, the value of
$\asmz=0.1176\,^{+0.0014}_{-0.0016}$ is obtained, in good agreement with the world average. It is the most precise result for \asmz from jet cross sections to date. At the same time, this analysis demonstrates that improved PDFs can be extracted by using these data.
Furthermore, by extracting the value of \as in different intervals of jet \pt,
the running of \as is tested at NNLO in QCD up to energy scales of 1.6\TeV.
The observed running of \as
is in agreement  with the prediction of the QCD renormalization group equation.

\begin{acknowledgments}
We congratulate our colleagues in the CERN accelerator departments for the excellent performance of the LHC and thank the technical and administrative staffs at CERN and at other CMS institutes for their contributions to the success of the CMS effort. In addition, we gratefully acknowledge the computing centres and personnel of the Worldwide LHC Computing Grid and other centres for delivering so effectively the computing infrastructure essential to our analyses. Finally, we acknowledge the enduring support for the construction and operation of the LHC, the CMS detector, and the supporting computing infrastructure provided by the following funding agencies: SC (Armenia), BMBWF and FWF (Austria); FNRS and FWO (Belgium); CNPq, CAPES, FAPERJ, FAPERGS, and FAPESP (Brazil); MES and BNSF (Bulgaria); CERN; CAS, MoST, and NSFC (China); MINCIENCIAS (Colombia); MSES and CSF (Croatia); RIF (Cyprus); SENESCYT (Ecuador); ERC PRG, RVTT3 and MoER TK202 (Estonia); Academy of Finland, MEC, and HIP (Finland); CEA and CNRS/IN2P3 (France); SRNSF (Georgia); BMBF, DFG, and HGF (Germany); GSRI (Greece); NKFIH (Hungary); DAE and DST (India); IPM (Iran); SFI (Ireland); INFN (Italy); MSIP and NRF (Republic of Korea); MES (Latvia); LMTLT (Lithuania); MOE and UM (Malaysia); BUAP, CINVESTAV, CONACYT, LNS, SEP, and UASLP-FAI (Mexico); MOS (Montenegro); MBIE (New Zealand); PAEC (Pakistan); MES and NSC (Poland); FCT (Portugal); MESTD (Serbia); MCIN/AEI and PCTI (Spain); MOSTR (Sri Lanka); Swiss Funding Agencies (Switzerland); MST (Taipei); MHESI and NSTDA (Thailand); TUBITAK and TENMAK (Turkey); NASU (Ukraine); STFC (United Kingdom); DOE and NSF (USA).
    
\hyphenation{Rachada-pisek} Individuals have received support from the Marie-Curie programme and the European Research Council and Horizon 2020 Grant, contract Nos.\ 675440, 724704, 752730, 758316, 765710, 824093, 101115353, 101002207, and COST Action CA16108 (European Union); the Leventis Foundation; the Alfred P.\ Sloan Foundation; the Alexander von Humboldt Foundation; the Science Committee, project no. 22rl-037 (Armenia); the Belgian Federal Science Policy Office; the Fonds pour la Formation \`a la Recherche dans l'Industrie et dans l'Agriculture (FRIA-Belgium); the F.R.S.-FNRS and FWO (Belgium) under the ``Excellence of Science -- EOS" -- be.h project n.\ 30820817; the Beijing Municipal Science \& Technology Commission, No. Z191100007219010 and Fundamental Research Funds for the Central Universities (China); the Ministry of Education, Youth and Sports (MEYS) of the Czech Republic; the Shota Rustaveli National Science Foundation, grant FR-22-985 (Georgia); the Deutsche Forschungsgemeinschaft (DFG), among others, under Germany's Excellence Strategy -- EXC 2121 ``Quantum Universe" -- 390833306, and under project number 400140256 - GRK2497; the Hellenic Foundation for Research and Innovation (HFRI), Project Number 2288 (Greece); the Hungarian Academy of Sciences, the New National Excellence Program - \'UNKP, the NKFIH research grants K 131991, K 133046, K 138136, K 143460, K 143477, K 146913, K 146914, K 147048, 2020-2.2.1-ED-2021-00181, TKP2021-NKTA-64, and 2021-4.1.2-NEMZ\_KI-2024-00036 (Hungary); the Council of Science and Industrial Research, India; ICSC -- National Research Centre for High Performance Computing, Big Data and Quantum Computing and FAIR -- Future Artificial Intelligence Research, funded by the NextGenerationEU program (Italy); the Latvian Council of Science; the Ministry of Education and Science, project no. 2022/WK/14, and the National Science Center, contracts Opus 2021/41/B/ST2/01369 and 2021/43/B/ST2/01552 (Poland); the Funda\c{c}\~ao para a Ci\^encia e a Tecnologia, grant CEECIND/01334/2018 (Portugal); the National Priorities Research Program by Qatar National Research Fund; MCIN/AEI/10.13039/501100011033, ERDF ``a way of making Europe", and the Programa Estatal de Fomento de la Investigaci{\'o}n Cient{\'i}fica y T{\'e}cnica de Excelencia Mar\'{\i}a de Maeztu, grant MDM-2017-0765 and Programa Severo Ochoa del Principado de Asturias (Spain); the Chulalongkorn Academic into Its 2nd Century Project Advancement Project, and the National Science, Research and Innovation Fund via the Program Management Unit for Human Resources \& Institutional Development, Research and Innovation, grant B39G670016 (Thailand); the Kavli Foundation; the Nvidia Corporation; the SuperMicro Corporation; the Welch Foundation, contract C-1845; and the Weston Havens Foundation (USA).
\end{acknowledgments}

\bibliography{auto_generated}
\appendix
\numberwithin{figure}{section}
\section{JES correlation of inclusive jet measurements}
\label{sec:JES}

\begin{figure*}[!ht]
\centering
\includegraphics[width=0.9\textwidth]{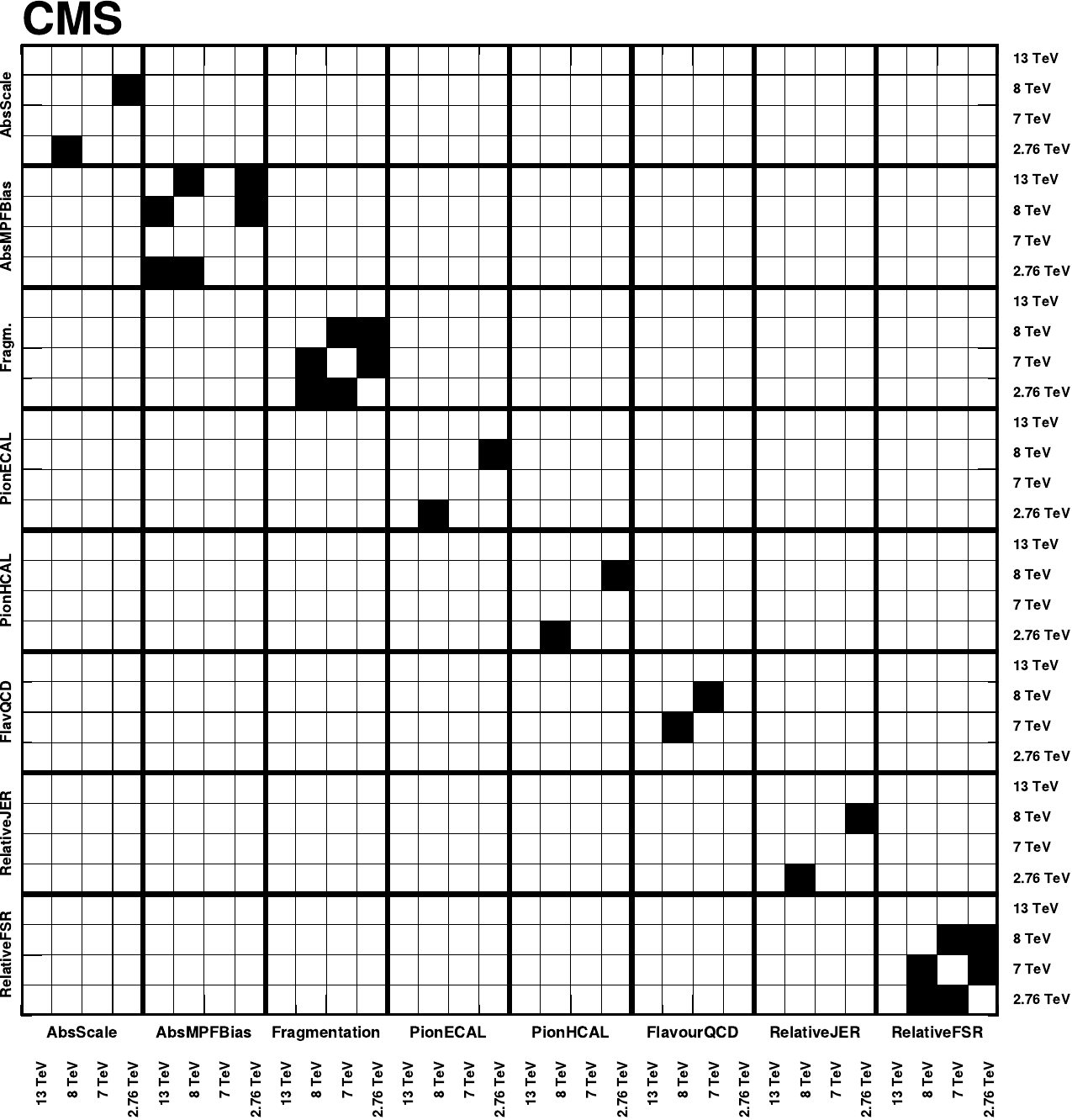}
\caption{Sketch of the treatment of JES sources of uncertainty among the measurements at different \sqrts. A filled black box indicates that the corresponding two uncertainty sources are treated as correlated.}
\label{fig:JEC}
\end{figure*}

The JES correlation scheme as illustrated in Fig.~\ref{fig:JEC} has been applied in the QCD analysis of the CMS inclusive jet measurements at $\sqrts=2.76$, 7, 8, and 13\TeV. A filled black box indicates that the corresponding two measurements are treated as fully correlated for the specific source of uncertainty.

\cleardoublepage \section{The CMS Collaboration \label{app:collab}}\begin{sloppypar}\hyphenpenalty=5000\widowpenalty=500\clubpenalty=5000
\cmsinstitute{Yerevan Physics Institute, Yerevan, Armenia}
{\tolerance=6000
V.~Chekhovsky, A.~Hayrapetyan, V.~Makarenko\cmsorcid{0000-0002-8406-8605}, A.~Tumasyan\cmsAuthorMark{1}\cmsorcid{0009-0000-0684-6742}
\par}
\cmsinstitute{Institut f\"{u}r Hochenergiephysik, Vienna, Austria}
{\tolerance=6000
W.~Adam\cmsorcid{0000-0001-9099-4341}, J.W.~Andrejkovic, L.~Benato\cmsorcid{0000-0001-5135-7489}, T.~Bergauer\cmsorcid{0000-0002-5786-0293}, S.~Chatterjee\cmsorcid{0000-0003-2660-0349}, K.~Damanakis\cmsorcid{0000-0001-5389-2872}, M.~Dragicevic\cmsorcid{0000-0003-1967-6783}, P.S.~Hussain\cmsorcid{0000-0002-4825-5278}, M.~Jeitler\cmsAuthorMark{2}\cmsorcid{0000-0002-5141-9560}, N.~Krammer\cmsorcid{0000-0002-0548-0985}, A.~Li\cmsorcid{0000-0002-4547-116X}, D.~Liko\cmsorcid{0000-0002-3380-473X}, I.~Mikulec\cmsorcid{0000-0003-0385-2746}, J.~Schieck\cmsAuthorMark{2}\cmsorcid{0000-0002-1058-8093}, R.~Sch\"{o}fbeck\cmsAuthorMark{2}\cmsorcid{0000-0002-2332-8784}, D.~Schwarz\cmsorcid{0000-0002-3821-7331}, M.~Sonawane\cmsorcid{0000-0003-0510-7010}, W.~Waltenberger\cmsorcid{0000-0002-6215-7228}, C.-E.~Wulz\cmsAuthorMark{2}\cmsorcid{0000-0001-9226-5812}
% ...
\par}
\vskip\cmsinstskip
\dag:~Deceased\\
$^{1}$Also at Yerevan State University, Yerevan, Armenia\\
$^{2}$Also at TU Wien, Vienna, Austria\\
$^{3}$Also at Ghent University, Ghent, Belgium\\
$^{4}$Also at Universidade do Estado do Rio de Janeiro, Rio de Janeiro, Brazil\\
$^{5}$Also at FACAMP - Faculdades de Campinas, Sao Paulo, Brazil\\
$^{6}$Also at Universidade Estadual de Campinas, Campinas, Brazil\\
$^{7}$Also at Federal University of Rio Grande do Sul, Porto Alegre, Brazil\\
$^{8}$Also at University of Chinese Academy of Sciences, Beijing, China\\
$^{9}$Also at China Center of Advanced Science and Technology, Beijing, China\\
$^{10}$Also at University of Chinese Academy of Sciences, Beijing, China\\
$^{11}$Also at China Spallation Neutron Source, Guangdong, China\\
$^{12}$Now at Henan Normal University, Xinxiang, China\\
$^{13}$Also at University of Shanghai for Science and Technology, Shanghai, China\\
$^{14}$Now at The University of Iowa, Iowa City, Iowa, USA\\
$^{15}$Also at an institute formerly covered by a cooperation agreement with CERN\\
$^{16}$Also at Cairo University, Cairo, Egypt\\
$^{17}$Also at Suez University, Suez, Egypt\\
$^{18}$Now at British University in Egypt, Cairo, Egypt\\
$^{19}$Also at Purdue University, West Lafayette, Indiana, USA\\
$^{20}$Also at Universit\'{e} de Haute Alsace, Mulhouse, France\\
$^{21}$Also at Istinye University, Istanbul, Turkey\\
$^{22}$Also at an international laboratory covered by a cooperation agreement with CERN\\
$^{23}$Also at The University of the State of Amazonas, Manaus, Brazil\\
$^{24}$Also at University of Hamburg, Hamburg, Germany\\
$^{25}$Also at RWTH Aachen University, III. Physikalisches Institut A, Aachen, Germany\\
$^{26}$Also at Bergische University Wuppertal (BUW), Wuppertal, Germany\\
$^{27}$Also at Brandenburg University of Technology, Cottbus, Germany\\
$^{28}$Also at Forschungszentrum J\"{u}lich, Juelich, Germany\\
$^{29}$Also at CERN, European Organization for Nuclear Research, Geneva, Switzerland\\
$^{30}$Also at HUN-REN ATOMKI - Institute of Nuclear Research, Debrecen, Hungary\\
$^{31}$Now at Universitatea Babes-Bolyai - Facultatea de Fizica, Cluj-Napoca, Romania\\
$^{32}$Also at MTA-ELTE Lend\"{u}let CMS Particle and Nuclear Physics Group, E\"{o}tv\"{o}s Lor\'{a}nd University, Budapest, Hungary\\
$^{33}$Also at HUN-REN Wigner Research Centre for Physics, Budapest, Hungary\\
$^{34}$Also at Physics Department, Faculty of Science, Assiut University, Assiut, Egypt\\
$^{35}$Also at Punjab Agricultural University, Ludhiana, India\\
$^{36}$Also at University of Visva-Bharati, Santiniketan, India\\
$^{37}$Also at Indian Institute of Science (IISc), Bangalore, India\\
$^{38}$Also at Amity University Uttar Pradesh, Noida, India\\
$^{39}$Also at IIT Bhubaneswar, Bhubaneswar, India\\
$^{40}$Also at Institute of Physics, Bhubaneswar, India\\
$^{41}$Also at University of Hyderabad, Hyderabad, India\\
$^{42}$Also at Deutsches Elektronen-Synchrotron, Hamburg, Germany\\
$^{43}$Also at Isfahan University of Technology, Isfahan, Iran\\
$^{44}$Also at Sharif University of Technology, Tehran, Iran\\
$^{45}$Also at Department of Physics, University of Science and Technology of Mazandaran, Behshahr, Iran\\
$^{46}$Also at Department of Physics, Faculty of Science, Arak University, ARAK, Iran\\
$^{47}$Also at Helwan University, Cairo, Egypt\\
$^{48}$Also at Italian National Agency for New Technologies, Energy and Sustainable Economic Development, Bologna, Italy\\
$^{49}$Also at Centro Siciliano di Fisica Nucleare e di Struttura Della Materia, Catania, Italy\\
$^{50}$Also at Universit\`{a} degli Studi Guglielmo Marconi, Roma, Italy\\
$^{51}$Also at Scuola Superiore Meridionale, Universit\`{a} di Napoli 'Federico II', Napoli, Italy\\
$^{52}$Also at Fermi National Accelerator Laboratory, Batavia, Illinois, USA\\
$^{53}$Also at Lulea University of Technology, Lulea, Sweden\\
$^{54}$Also at Consiglio Nazionale delle Ricerche - Istituto Officina dei Materiali, Perugia, Italy\\
$^{55}$Also at Institut de Physique des 2 Infinis de Lyon (IP2I ), Villeurbanne, France\\
$^{56}$Also at Department of Applied Physics, Faculty of Science and Technology, Universiti Kebangsaan Malaysia, Bangi, Malaysia\\
$^{57}$Also at Consejo Nacional de Ciencia y Tecnolog\'{i}a, Mexico City, Mexico\\
$^{58}$Also at Trincomalee Campus, Eastern University, Sri Lanka, Nilaveli, Sri Lanka\\
$^{59}$Also at Saegis Campus, Nugegoda, Sri Lanka\\
$^{60}$Also at National and Kapodistrian University of Athens, Athens, Greece\\
$^{61}$Also at Ecole Polytechnique F\'{e}d\'{e}rale Lausanne, Lausanne, Switzerland\\
$^{62}$Also at Universit\"{a}t Z\"{u}rich, Zurich, Switzerland\\
$^{63}$Also at Stefan Meyer Institute for Subatomic Physics, Vienna, Austria\\
$^{64}$Also at Laboratoire d'Annecy-le-Vieux de Physique des Particules, IN2P3-CNRS, Annecy-le-Vieux, France\\
$^{65}$Also at Near East University, Research Center of Experimental Health Science, Mersin, Turkey\\
$^{66}$Also at Konya Technical University, Konya, Turkey\\
$^{67}$Also at Izmir Bakircay University, Izmir, Turkey\\
$^{68}$Also at Adiyaman University, Adiyaman, Turkey\\
$^{69}$Also at Bozok Universitetesi Rekt\"{o}rl\"{u}g\"{u}, Yozgat, Turkey\\
$^{70}$Also at Marmara University, Istanbul, Turkey\\
$^{71}$Also at Milli Savunma University, Istanbul, Turkey\\
$^{72}$Also at Kafkas University, Kars, Turkey\\
$^{73}$Now at Istanbul Okan University, Istanbul, Turkey\\
$^{74}$Also at Hacettepe University, Ankara, Turkey\\
$^{75}$Also at Erzincan Binali Yildirim University, Erzincan, Turkey\\
$^{76}$Also at Istanbul University -  Cerrahpasa, Faculty of Engineering, Istanbul, Turkey\\
$^{77}$Also at Yildiz Technical University, Istanbul, Turkey\\
$^{78}$Also at School of Physics and Astronomy, University of Southampton, Southampton, United Kingdom\\
$^{79}$Also at IPPP Durham University, Durham, United Kingdom\\
$^{80}$Also at Monash University, Faculty of Science, Clayton, Australia\\
$^{81}$Also at Universit\`{a} di Torino, Torino, Italy\\
$^{82}$Also at Bethel University, St. Paul, Minnesota, USA\\
$^{83}$Also at Karamano\u {g}lu Mehmetbey University, Karaman, Turkey\\
$^{84}$Also at California Institute of Technology, Pasadena, California, USA\\
$^{85}$Also at United States Naval Academy, Annapolis, Maryland, USA\\
$^{86}$Also at Ain Shams University, Cairo, Egypt\\
$^{87}$Also at Bingol University, Bingol, Turkey\\
$^{88}$Also at Georgian Technical University, Tbilisi, Georgia\\
$^{89}$Also at Sinop University, Sinop, Turkey\\
$^{90}$Also at Erciyes University, Kayseri, Turkey\\
$^{91}$Also at Horia Hulubei National Institute of Physics and Nuclear Engineering (IFIN-HH), Bucharest, Romania\\
$^{92}$Now at another institute formerly covered by a cooperation agreement with CERN\\
$^{93}$Also at Texas A\&M University at Qatar, Doha, Qatar\\
$^{94}$Also at Kyungpook National University, Daegu, Korea\\
$^{95}$Also at Yerevan Physics Institute, Yerevan, Armenia\\
$^{96}$Also at another international laboratory covered by a cooperation agreement with CERN\\
$^{97}$Also at Imperial College, London, United Kingdom\\
$^{98}$Also at Institute of Nuclear Physics of the Uzbekistan Academy of Sciences, Tashkent, Uzbekistan\\
$^{99}$Also at another institute formerly covered by a cooperation agreement with CERN\\
\end{sloppypar}
\end{document}